\documentclass{article}

\usepackage{arxiv}
\usepackage[utf8]{inputenc} 
\usepackage[T1]{fontenc} 
\usepackage{hyperref} 
\usepackage{url} 
\usepackage{booktabs} 
\usepackage{amsfonts} 
\usepackage{nicefrac} 
\usepackage{physics}
\usepackage{blkarray} 
\usepackage{amsmath}
\usepackage{subcaption}
\usepackage{graphicx}
\graphicspath{{Data/}}
\title{Computational Chemistry on Quantum Computers}

\author{
 Jerimiah B.~Wright\thanks{All simulations studied used Qumquat, an experimental high-level quantum programming language. https://github.com/patrickrall/Qumquat} \\
 Department of Chemistry\\
 University of Texas \\
 Austin, TX 78705 \\
 \texttt{Jerimiah.Wright@utexas.edu} \\
}

\begin{document}
\maketitle {}
\begin{abstract}
 The purpose of this experiment was to use the known analytical techniques to study the creation, simulation, and measurements of molecular Hamiltonians. The techniques used consisted of the Linear Combination of Atomic Orbitals (LCAO), the Linear Combination of Unitaries (LCU), and the Phase Estimation Algorithm (PEA). The molecules studied were $H_2$ with and without spin, as well as $He_2$ without spin. Hamiltonians were created under the LCAO basis, and reconstructed using the Jordan-Winger transform in order to create a linear combination of Pauli spin operators. The lengths of each molecular Hamiltonian greatly increased from the $H_2$ without spin, to $He_2$. This resulted in a reduced ability to simulate the Hamiltonians under ideal conditions. Thus, only low orders of l = 1 and l = 2 were used when expanding the Hamiltonian in accordance to the LCU method of simulation. The resulting Hamiltonians were measured using PEA, and plotted against function of $\frac{2\pi(K)}{N}$ and the probability distribution of each register. The resolution of the graph was dependent on the amount of registers, N, being used. However, the reduction of order hardly changed the image of the $H_2$ graphs.
\end{abstract}

\section{Introduction}
\subsection{Computational Chemistry}

Computational chemistry is the combination and application of chemical, mathematical and computing skills, used to solve chemical problems. These problems range from the study of molecular structure, electronic charge density distributions, reactivity, and many more. These computational results are usually used to complement experimental results, however as they can also shed light on unobserved chemical phenomena, they are also used to design new drugs and materials. 

The Schroëdinger equation is one of the starting points for most quantum chemical calculations. It can be used to predict or calculate the energy of atoms as particles move around. However, for many body systems, the Schroëdinger equation cannot be solved analytically. To this end, many chemist use different mathematical methods to approximate solutions to the equation. These methods range from simplified forms of the first-principles equations that are easier or faster to solve, to methods that limit the size of the system such as periodic boundary conditions. Because approximation methods are required to achieve solutions to the Schroëdinger equation, the goal of computational chemist has been to minimize the residual error of the approximations while keeping the calculations tractable.

Performing any kind of chemical computation requires a method of representing the system. These methods are known as basis sets, and are typically comprised of a set of functions that are used to represent a wave function. One of the more common basis sets is composed of the Linear Combination of Atomic Orbitals. LCAO is a quantum superposition of atomic orbitals and a technique for calculating molecular orbitals in quantum chemistry. Assuming the molecular orbital $\phi_i$ is equal to the number of atomic orbitals included in the linear expansion, the expression $\phi_i = \sum_{j} c_{ij} \chi_j $ can be formed where where 
 $\phi _{i}$ is a molecular orbital represented as the sum of n atomic orbitals 
 $\chi _{r} $, each multiplied by a corresponding coefficient 
 $\ c_{ij}$, and j represents which atomic orbital is combined in the term. This method can be incredibly useful when trying to describe molecular orbitals because of its simplicity.

\subsection{The Molecular Hamiltonian}

A Hamiltonian operator, \^{H}, is an operator corresponding to the sum of the potential and kinetic energies of all particles in a system, $\ket{\psi}$. Using this operator, the Schroëdinger equation can be expressed as $\hat{H} \ket{\psi} = E_a\ket{\psi}$. In the case of molecules, this Hamiltonian is known as the Molecular Hamiltonian. The most common mathematical approximation is the Bohrn and Oppenheimer approximation in which nuclear kinetic energy terms within the Hamiltonian are neglected. This results in a Hamiltonian that only considers the kinetic energies of electrons and the Coulomb interactions between them. This new Hamiltonian is known as the Electronic Hamiltonian, and is shown in its second quantized form below. 

\begin{equation}
 \hat{H} = \sum_{i,j} h_{i,j} a^\dagger_ia_j + \dfrac{1}{2}\sum_{i,j,k,l} h_{i,j,k,l} a^\dagger_ia^\dagger_ka_ja_l
 \label{ Eq. 1}
\end{equation}

The Electronic Hamiltonian's second quantized form tracks each orbital and stores whether there is an electron present in each of them. In order to do this annihilation and creation operators are used as they can act on electronic states. The annihilation operator is denoted $a$, and it lowers the number of particles in a given state by one. The creation operator is denoted $a^{\dagger}$ and it increases the number of particles in a given state by one, as well as being the complex conjugate of the annihilation operator. The coefficients $h_{ij}$ and $h_{ijkl}$ are one-and two-electron overlap integrals which can be computed classically. 

\label{sec:headings}
\subsection{Quantum Representation}

Computational chemist often use computers to solve chemical problems like molecular structure. Classical software packages are often limited as they must store an electrons wave function of each particle as a probability distribution. This becomes a problem as the memory required to store the wave function in the form of bits scales exponentially with the number of electrons in the molecule. With the introduction of quantum bits, or qubits, a particles wave function can be more efficiently stored. This is because qubits, as the name implies, are quantum allowing them to be described as wave functions themselves. 

In practical applications, the qubits used for computation must be distinguishable. This becomes an issue when they are used to represent the electrons of molecular systems as the electrons are indistinguishable particles. In order to circumvent this issue, the Jordan-Winger transform is used to express the annihilation operator $a_j$, and the creation operator $a^\dagger_j$, in terms of the Pauli spin operators X,Y,Z and the identity I, that correspond to the algebra of distinguishable $\nicefrac{1}{2}$ spin particles. The Jordan-Winger transform is given by the equation below. 

\begin{equation}
 \begin{split}
 a_j \equiv I ^{\otimes n - j - 1} \otimes \hat{Q}^- \otimes Z^{\otimes j} \\ a^\dagger_j \equiv I ^{\otimes n - j - 1} \otimes \hat{Q}^+ \otimes Z^{\otimes j}
 \end{split}
\end{equation}
\begin{center}
  *Note that $\hat{Q}^- = \ket{0}\bra{1} = \frac{1}{2} (X + iY) $ and $\hat{Q}^+ = \ket{1}\bra{0} = \frac{1}{2} (X - iY) $.
\end{center} 
Once the Hamiltonian has been expressed as a sum of elementary Pauli spin operators the dynamics can then be compiled into fundamental gate operations using a host of well-known techniques. One such technique is the Linear Combination of Unitaries, or LCU. 

The LCU algorithm is used to solve the Hamiltonian Simulation problem, where the goal is construct a unitary operation U that follows the equation shown below.
\begin{equation}
  \norm{U - e^{-iHt }} \leq \epsilon 
\end{equation}
\begin{center}
  *Note that $\epsilon$ represents the error parameter.
\end{center}

 Suppose there is unitary that can can be written as a linear combination of efficiently implementable unitary matrices U$_i$ , i.e., $V = \sum_{i} a_i U_i$. This unitary U then maps $\ket{i}\ket{\psi}$ to $\ket{i} U_i \ket{\psi}$. This means that if the Hamiltonian itself is a linear combination of unitaries where U$_i$ is unitary for all i, then the map $e^{iHt}$, is also a linear combination of unitaries.

\begin{equation}
 e^{-iHt} = \sum_{l=0}^\infty \frac{1}{l!}(-iHt)^l = \sum_{l=0}^\infty \frac{1}{l!}(-i (\sum_{i} a_i U_i t)^l
\end{equation}

When expanded out, each term is a scalar multiplied by a product of unitary operators. Although this is an infinite sum, a good approximation can be obtained by truncating the sum. Using this idea, a similar method can be applied to simulate sparse Hamiltonians that are not restricted to being Hermitian. The new operator V is defined as $V : = e^{{-iH}/{m}} = e^{{-i\sum_{j} U_j}/{m}}$. It is important to note that the coefficient $a_i$ can be absorbed int the unitary $U_i$ without any loss in relative phase. Expanding the definition of the matrix exponential, and truncating the sum after l = k results in the matrix $\Tilde{V}$.

\begin{equation}
 \Tilde{V} := \sum_{l=0}^k \dfrac{1}{l! m^l} (-i \sum_{j} U_j)^l
 \label{Eq. 5}
\end{equation}

\section{Experimental}
\subsection{Creating the Hamiltonian}

The molecules of interest in this experiment were diatomic hydrogen and helium.
A unique form of the electronic Hamiltonian (\ref{ Eq. 1}) was used to simulate the quantum chemistry of the molecules. This equation is shown below. 

\begin{equation}
 \hat{H} = \sum_{i,j} h_{ij} b^\dagger_ib_j
 \label{Eq. 6}
\end{equation}

The Hamiltonian was created using the LCAO method to rewrite the molecular orbitals ,$b^\dagger b$, in terms of separable atomic orbitals. By treating each atomic orbital as a vector, ie. $\ket{1s_A}$, a matrix of inner products, $\bra{1s_A}\ket{1s_B}$ could be produced. This matrix is known as the overlap matrix. 
\begin{figure}[h]
 \centering
 \includegraphics[]{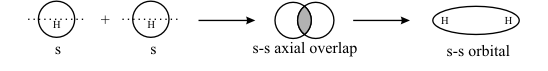}
 \caption{The overlap between the 1s orbitals of Hydrogen}
 \label{fig:fig1}
\end{figure}

In order to write the the molecular orbitals as separable atomic orbitals, the Gram-Schmidt process was used to create an orthonormal set of virtual orbitals that do not overlap. These virtual orbitals were substituted into equation [\ref{Eq. 6}] in order to produce a separable molecular Hamiltonian. The Jordan-Winger transform was performed on the new molecular Hamiltonians, resulting in their decomposition into Pauli spin operators.

\subsection{Hamiltonian Simulation}
Once the Hamiltonian has been expressed as a sum of elementary Pauli spin, the LCU method could be used to simulate Hamiltonian dynamics. First, A unitary matrix, A, was created to map $\ket{0}$ to $\frac{1}{\sqrt{a}} \sum_{i} \sqrt{a_i}\ket{i}$. This was done using the postselect feature in Qumquat, the Quantum Machine Learning and Quantum Algorithms Toolkit. The Palui string Hamiltonians were then expanded in accordance to equation [\ref{Eq. 5}]. The hydrogen molecule without spin was expanded with orders of l = 1 and l = 2, while the hydrogen molecule with spin was expanded only with an order of l = 2. The helium molecule without spin was expanded with an order of l = 1. The resulting expand Pauli strings were applied to the quantum circuit, conversting the eigenenergy into relative phases. The relative phases can then be measred using the Phase Estimation algorithm. 

\subsection{Measuring the Hamiltonian}
The phase estimation algorithm, or PEA, was used to obtain the molecular energies from the time evolution of the molecular Hamiltonian described in section 2.2. Given a unitary matrix U and a quantum state $\ket{\psi}$ such that U$\ket{\psi} = e^{2\pi i \theta}$, the algorithm estimates the value of $\theta$ with high probability with additive error $\epsilon$. In the case of calculating molecular energies, the unitary matrix U is the molecular Hamiltonian produced after simulating its dynamics. The algorithm begins by setting up two registers where the upper n qubits comprise the first register, and the lower m qubits are the second register. A super position of the initial state is created by applying n-bit Hadamard gate operations on the first register, resulting in the state below.

\begin{equation}
 \frac{1}{2^{\nicefrac{n}{2}}}(\ket{0} + \ket{1})^{\otimes n}
\end{equation} 

A controlled-U gate is applied to the second register only if its corresponding control qubit is $\ket{1}$. Applying n many controlled-U gate operations to the phase of an eigenstate of the Hamiltonian evolving dependent on a register qubit, i.e. $\ket{0}\ket{\psi_n}+ e^{-iHt}\ket{1}\ket{\psi_n} = \ket{0}\ket{\psi_n}+ e^{-iE_nt}\ket{1}\ket{\psi_n}$, allows the state of the first register to be described by the following equation.

\begin{equation}
 \frac{1}{2^{\nicefrac{n}{2}}} \sum_{k=0}^{N} e^{-iE_nt} 
\end{equation}
\begin{center}
  * Note that N = $2^n - 1$
\end{center}

When $E_n = 2\pi(\theta - K)/t$ where $0 \leq \theta \leq 1$ and K is an unobservable integer, the unknown eigenvalue becomes encoded in the relative phase of the register quantum state as $\ket{0} + e^{-2\pi i(\theta - K)}\ket{1}$. The $\theta$ can then be estimated by applying an inverse quantum Fourier transform, $\frac{1}{2^{\nicefrac{n}{2}}} \sum_{k=0}^{N} e^{\frac{-2\pi i n \theta}{N}}$, and performing a measurement in the computational basis on the first register.

\section{Results}

\begin{figure}[h]
 \centering
 \centering
 \[
 H_2 (no-spin) = 
 \begin{blockarray}{ccc}
 1S_A & 1S_B\\
 \begin{block}{[cc]c}
 1 & S & 1S_A\\
 S & 1 & 1S_B\\
 \end{block}
 \end{blockarray} 
 \]
 \[ 
 H_2 (spin) = \begin{blockarray}{ccccc}
 1s_{A\uparrow} & 1s_{A\downarrow} & 1s_{B\uparrow} & 1s_{B\downarrow} \\
 \begin{block}{[cccc]c}
 1 & S & 0 & 0 & 1s_{A\uparrow} \\
 S & 1 & 0 & 0 & 1s_{A\downarrow} \\
 0 & 0 & 1 & S & 1s_{B\uparrow} \\
 0 & 0 & S & 1 & 1s_{B\downarrow} \\
 \end{block}
 \end{blockarray}
 \]

 \[
 He_2(no-spin) =
 \begin{blockarray}{ccccc}
 1s_A &1s_B & 2s_A & 2s_B \\
 \begin{block}{[cccc]c}
 1 & S & 0 & S_1 & 1s_A \\
 S & 1 & S_1 & 0 & 1s_B \\
 0 & S_1 & 1 & S_2 & 2s_A \\
 S_1 & 0 & S_2 & 1 & 2s_B \\
 \end{block}
 \end{blockarray}
 \]
 \caption{Overlap Matrices for $H_2$ without spin, $H_2$ with spin, and $He_2$ without spin}
 \label{fig:fig2}
\end{figure}

The LCAO approach was used in order to create a basis set in which the molecular Hamiltonian was a linear combination of atomic orbitals. Writing the Hamiltonian in this format aided in state preparation for Hamiltonian Simulation.The molecules were written in terms of their overlap matrix (Figure 2), and their atomic orbitals were separated using the Gram-Schmit process. The original orbitals were then written in terms of these new virtual orbitals, in order to be used in equation [\ref{Eq. 6}]. Performing the Jordan-Winger transform results in the Pauli decompositions of molecular Hamiltonians where each varied in length increasing from $H_2$ to $He_2$(Table 1). The amount of gate operations greatly increased the time it took to simulate and measure the results of each Hamiltoniain. 

\begin{table}[h]
 \caption{Table of Molecular Hamiltonians comprised of sums of Pauli operators}
 \centering
 \begin{tabular}{cccccc}
 \toprule
 \cmidrule(r){1-6}
 $H_2$ (no-spin) && $H_2$ (spin) && $He_2$ (no-spin) \\
 \midrule
 1.5686986355290005 & II	& 3.137397271058001 & IIII	& 22.452642369057198 & IIII \\
 -1.2843493177645002 & ZI &	-1.2843493177645002 & ZIII & -1.511819721289001 & ZIII\\
 -0.2843493177645001 & IZ &	-0.2843493177645001 & IZII	&-1.3117878918224433& IZII\\
 -0.4722596673392581 & XY &	-1.2843493177645002 & IIZI	&-5.029764413362022 &IIZI\\
 0.4722596673392581 & YX &-0.2843493177645001 &IIIZ&	-14.599270342583733& IIIZ\\
 &&	-0.4722596673392581 & XYII&	-0.09446696524113857 &XYII \\
 &&	0.4722596673392581& YXII&	0.09446696524113857 &YXII \\
 &&	-0.4722596673392581& IIXY&	-0.9924233565701107 &XZYI\\
 &&	0.4722596673392581 &IIYX&	0.9924233565701107 & YZXI\\
 &&	&&	-1.1191380207487813& IXXI\\ 
 &&	&&	-1.1191380207487813& IYYI\\
 &&	&&	1.8223341943756635 &XZZY\\
 &&	&&	-1.8223341943756635 &YZZX\\
 &&	&&	3.026611588811937 &IXZX\\
 &&	&&	3.026611588811937 &IYZY\\
 &&	&&	-7.9506036387717955& IIXX\\
 &&	&&	-7.9506036387717955 & IIYY\\
 \bottomrule
 \end{tabular}
 \label{tab:table 1}
\end{table}

Hamiltonian Simulation was done using the LCU method. Since the generated Hamiltonians were already linear combinations of Pauli spin operators, promoting them into equation [\ref{Eq. 6}] was simple. However, because of the length of each Hamiltonian, only low orders of l could be studied, as the length would increase exponentially, drastically increasing the time it took to simulate the results.

The PEA algorithm was used to measure the eigenvalue of the of the Hamiltonian operator by encoding the relative phase of the register quantum state as $\ket{0} + e^{-i2\pi(\theta - K)/t}$ and applying an inverse quantum Fourier transform, them measuring on the first register. Plotting these results as a function of $\frac{2\pi(K)}{N}$ and probability distribution of each register, produced the graphs shown in figures 3 and 4 

\begin{figure}[h]
 \begin{subfigure}{.3\textwidth}
 \centering
 \includegraphics[width=.8\linewidth]{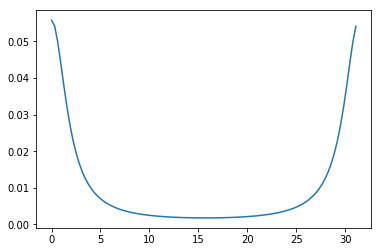}
 \caption{N = 100, l = 2}
 \label{fig:sfig1}
\end{subfigure}%
\begin{subfigure}{.3\textwidth}
 \centering
 \includegraphics[width=.8\linewidth]{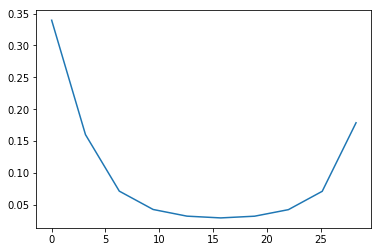}
 \caption{N = 10, l = 2}
 \label{fig:sfig2}
\end{subfigure}
\begin{subfigure}{.3\textwidth}
 \centering
 \includegraphics[width=.8\linewidth]{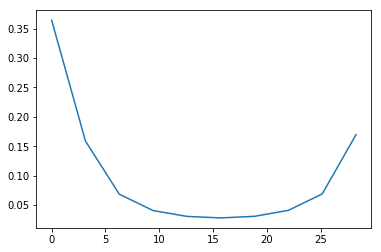}
 \caption{N = 10, l = 1}
 \label{fig:sfig3}
\end{subfigure}
\caption{Plots of the $H_2$ molecule without spin under different parameters.}
\label{fig:fig3}

\centering
\begin{subfigure}{.4\textwidth}
 \centering
 \includegraphics[width = .8\linewidth]{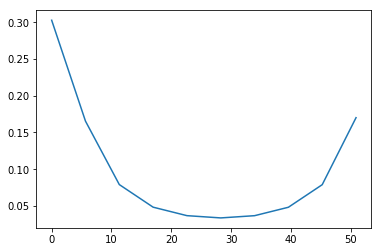}
 \caption{$H_2$ molecule with spin. N = 10, O = 2}
 \label{fig:sfig4}
\end{subfigure}
\begin{subfigure}{.4\textwidth}
 \centering
 \includegraphics[width = .8\linewidth]{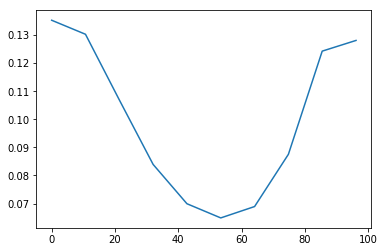}
 \caption{$He_2$ molecule with spin. N = 10, O = 1}
 \label{fig:sfig5}
\end{subfigure}
\caption{Plots of the $H_2$ and $He_2$ molecules}
\label{fig:fig4}
\end{figure}

\section{Conclusions}
The graphs shown in figures 3 and 4 demonstrate the results under different values of N and l, where N represents the range of register used, and l represents the order of the of the molecular Hamiltonian produced from equation [\ref{Eq. 5}]. As no error mitigation was done, these results should be studied qualitatively. As shown in figure \ref{fig:fig1}, decreasing the range of N drastically decreases the resolution of the graph, with the benefit of producing a graph much quicker. Comparing \ref{fig:sfig2} to \ref{fig:sfig3} shows the result of changing the order of the Hamiltonian expansion. For low values of l, the effect of changing the order seems to be fairly negligible. This is explored further in the appendix. Finally, the more complex Hamiltonians of $H_2$ with spin, and $He_2$ without spin are shown in figure 2a and 2b respectively. Despite Both graphs having low values of N and l, they have distinct shapes. However, $H_2$ molecule with spin shares a nearly identical appearance to its no-spin counter part. While these results must be interpreted qualitatively, they still help prove the methodology of producing the Hamiltonians as legitimate through reproducibility. If this method was flawless, it would be expected that the spin variant would have a higher quality graph despite having the same N and l values, due to a more complete description of the molecule. The $He_2$ molecule has a more unique shape indicating a clear difference between the two molecules. 

The purpose of this experiment was to use the known analytical techniques to study the creation, simulation, and measurements of molecular Hamiltonians. While this experiment was successful in implementing known techniques to perform electronic structure calculations, it was extremely limited due to the performance capabilities of classical computers. This greatly hindered the ability for quantitative results due to using low resolution graphs to identify the eigenvalues of each molecular Hamiltonian. 

\bibliographystyle{unsrt}

\end{document}